\documentclass[preprint2]{aastex}

\usepackage{graphicx}
\usepackage{amsmath, amssymb}
\usepackage{multirow}
\usepackage[version=3]{mhchem}

\usepackage{natbib}
\bibpunct{(}{)}{;}{a}{}{,}

\graphicspath{{./figures/}}

\newcommand{\reffig}[1]{Fig.~\ref{#1}}
\newcommand{\reftab}[1]{Table~\ref{#1}}

\newcommand{\OI}{O~\textsc{i}}
\newcommand{\CII}{C~\textsc{ii}}

\begin{document}

\title{Volatile depletion in the TW~Hydrae disk atmosphere}

\author{Fujun~Du$^{1}$, Edwin~A.~Bergin$^{1}$ \& Michiel R. Hogerheijde$^{2}$}

\affil{Department of Astronomy,
       University of Michigan,
       311 West Hall, 1085 S. University Ave,
       Ann Arbor, MI 48109, USA}
\affil{Leiden Observatory,
       Leiden University,
       Post Office Box 9513,
       2300 RA Leiden, Netherlands}
\email{fdu@umich.edu}


\begin{abstract}
An abundance decrease in carbon- and oxygen-bearing species relative to dust
has been frequently found in planet-forming disks, which can be attributed to
an overall reduction of gas mass.  However, in the case of TW~Hya, the only
disk with gas mass measured directly with HD rotational lines, the inferred gas
mass ($\lesssim$0.005 solar mass) is significantly below the directly measured
value ($\gtrsim$0.05 solar mass).  We show that this apparent conflict can be
resolved if the elemental abundances of carbon and oxygen are reduced in the
upper layers of the outer disk but are normal elsewhere (except for a possible
enhancement of their abundances in the inner disk).  The implication is that in
the outer disk, the main reservoir of the volatiles (CO, water, \ldots) resides
close to the midplane, locked up inside solid bodies that are too heavy to be
transported back to the atmosphere by turbulence.  An enhancement in the carbon
and oxygen abundances in the inner disk can be caused by inward migration of
these solid bodies.  This is consistent with estimates based on previous models
of dust grain dynamics.  Indirect measurements of the disk gas mass and disk
structure from species such as CO will thus be intertwined with the evolution
of dust grains, and possibly also with the formation of planetesimals.
\end{abstract}



\section{Introduction}

Observations and modeling of transitional disks have pointed to an
apparent lack of gas relative to their dust content, particularly at larger
distances from the star.  Early observations (see, e.g. \citealt{Skrutskie1991}
and \citealt{Yamashita1993}) show that the CO emission are much lower
than expected from dust emission and molecular cloud composition.
A recent Herschel/PACS survey finds that
transitional disks have weaker \OI{} 63~$\mu$m emission than their full disk
counterparts, even if they have similar continuum at 63~$\mu$m
\citep{Keane2014,Howard2013}.  One interpretation of this is that the gas mass
is reduced, or, equivalently in effect, that the transitional disks are less
flared than their full disk counterpart \citep{Keane2014}.  \citet{Aresu2014} find that with a normal
gas-to-dust ratio, models of the \OI{} 63~$\mu$m emission for disks in Taurus
tend to over-predict the line intensity.  Similarly, detailed
modeling of the prototypical transitional disk, TW~Hya, suggest a small gas
mass (0.5--5)$\times10^{-3}$~$M_\odot$ (and small gas-to-dust mass ratio)
\citep{Thi2010,Kamp2013,Williams2014}.

However, the gas mass in these studies are only inferred rather than directly
measured.  The main constituent of the gas is hydrogen, mostly in molecular
form, followed by helium, both of which defy direct detection in the bulk of
the disk, because their line excitation requires high temperature.  So the gas
mass is inferred from other agents, such as the commonly used CO lines or dust
continuum, in combination with radiative transfer and thermo-chemical models.
Fortunately, with Herschel, it is possible to measure the gas mass in a much
more direct way, by observing the rotational transition lines of HD, an
isotopologue of molecular hydrogen.  The translation of HD intensity to \ce{H2}
mass is more straightforward because there is no complex chemistry involved
(though the result depends on the temperature structure).  Direct measurement
of the gas mass of TW~Hya based on Herschel observation of HD lines gives mass
$\gtrsim$0.05~$M_\odot$ \citep{Bergin2013}, much higher than the values in some
of the previous studies.

To reconcile the two classes of seemingly disparate results on the gas mass of
TW~Hya, here we show that it is \emph{not} the overall gas mass --- the mass of
atomic and molecular hydrogen --- that is reduced, but rather it is the mass of
the ice-forming species, such as CO and H$_2$O, are reduced in the upper layers
of the outer disk.  It is these upper layers that are emissive and detectable
by astronomical observations.  The depletion of CO relative to hydrogen in DM~Tau and GG~Tau has been found by \citet{Dutrey1997} based on excitation arguments, and in TW~Hya by \citet{Favre2013}
based on direct comparison of the CO isotopologue
emission with the HD result.  Reduction of the water content
has been suggested by \citet{Bergin2010} and \citet[supporting
material]{Hogerheijde2011}.  Since CO and H$_2$O are the main bearers of carbon
and oxygen in the upper layers (below the photodissociated layer), the total
abundances of elemental carbon and oxygen are lowered.  This affect other
species chemically linked to them.  We also establish that this effect is
evident only in the outer disk (beyond the water and CO snow line), while in
the inner disk the opposite --- namely enhancement of the oxygen and carbon
abundances --- might be taking place.  A possible physical mechanism based on
dust grain settling and migration is provided at the end.

\section{Models}

\begin{figure}[htbp]
\centering
\includegraphics[width=\linewidth]{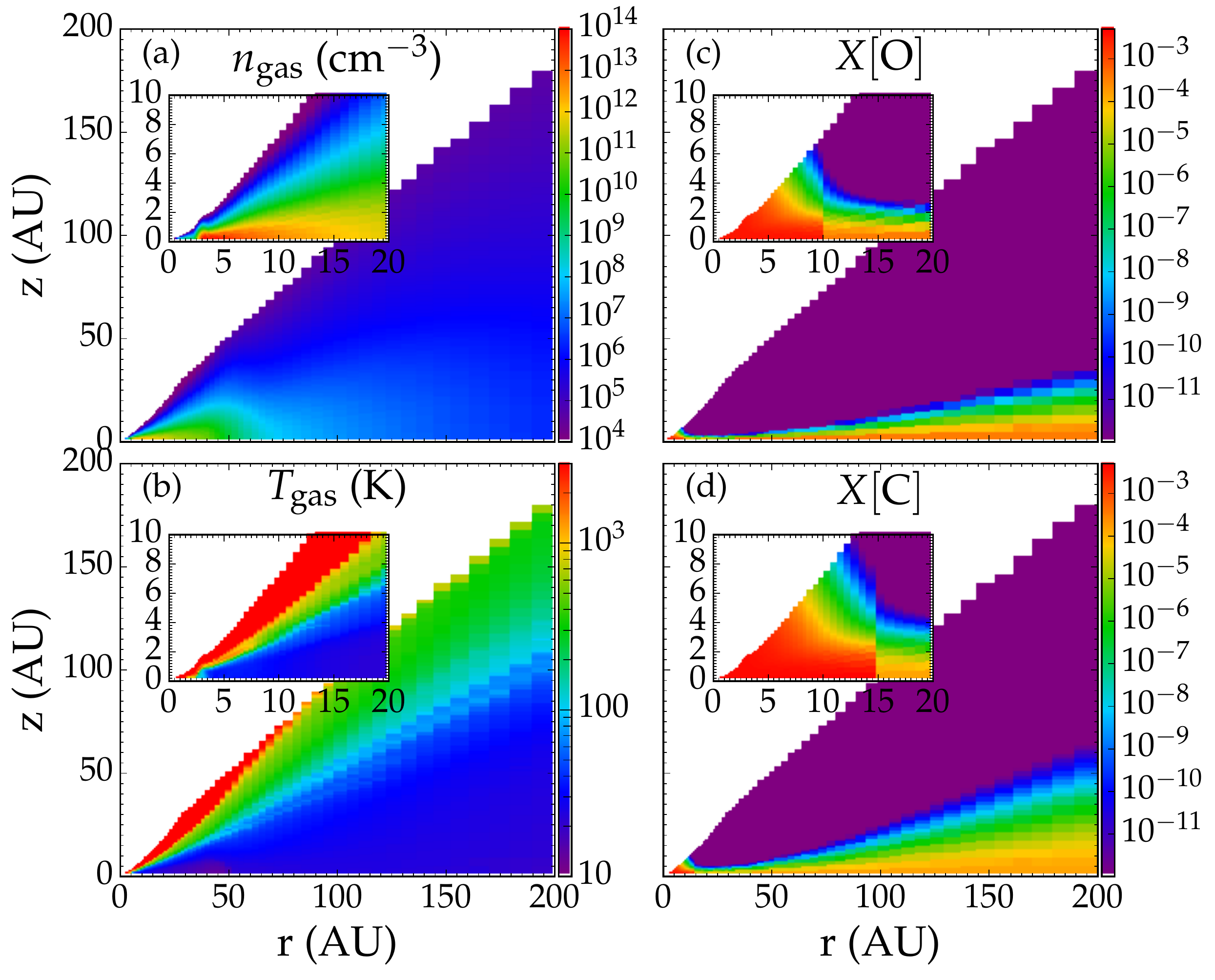}
\caption{Distribution of disk physical parameters used in the models.  (a): gas
density distribution in the two models.  (b): gas temperature distribution in
the model in which the oxygen and carbon abundances are changed; for the model
with full oxygen and carbon abundances, the gas temperature distribution is
slightly different due to changes in the heating and cooling rates of oxygen-
and carbon- bearing species.  (c) and (d): distribution of oxygen and carbon
abundances relative to hydrogen nuclei in the model in which their abundances
are changed.}
\label{fignTOC}
\end{figure}

We run two models for TW~Hya using a thermo-chemical and radiative-transfer
code \citep{Du2014}.  The system is evolved for 1~Myr.  Both models assume a disk with total gas mass
0.05~$M_\odot$ as constrained by the HD rotational lines \citep{Bergin2013}, a
distance of 51~pc \citep{Mamajek2005}, and an inclination angle of 7~degree
\citep{Qi2004}.  The dust mass is not very well constrained, so we use a
uniform gas-to-dust mass ratio of 100 for both models.  We assume three
components for the disk structure: an optically thin component inside
$\sim$3.5~AU, a component tapering at $\sim$50~AU \citep{Andrews2012,Menu2014},
and a component extended to $\sim$200~AU.  The mass ratio of the three
components are $10^{-4} : 20 : 1$. The outer two components have a surface
density profile $\Sigma \propto r^{-3/2}$.  The disk 3-D density and
temperature structure are determined based on vertical hydrostatic equilibrium,
calculated through iterative Monte Carlo simulation of dust radiative transfer,
which also yields a spectral energy distribution that matches observation.
The resulting 870~$\mu$m continuum map of TW~Hya resembles the observed one of
\citet{Andrews2012}, though we spend no effort to match the visibility profile.
The distribution of gas density and temperature is shown in panel (a) and (b)
of \reffig{fignTOC}.  The only difference between the two models lies in the
abundances of oxygen and carbon (relative to hydrogen).  The first model has
canonical ``low-metal'' ISM (interstellar medium) oxygen and carbon abundances
\citep{Garrod2008} in the whole disk, while the second model has reduced oxygen
and carbon abundances in the upper layers of the outer disk, and enhanced
oxygen and carbon abundances in the inner disk (see panel (c) and (d) of
\reffig{fignTOC}).  The degree of reduction for oxygen and carbon are based on
an analytical prescription, the exact form of which is manually adjusted for a
best fit.  Carbon is less reduced than oxygen in the outer disk, inspired by
the difference in the evaporation temperatures \citep{Oberg2011} of water and
the major bearers of carbon (CO and CO$_2$).  The enhancement in the inner disk
is simply implemented with a step function.
This empirical approach is particularly tailored to match the observational data.
Future self-consistent study is needed.

\section{Results}

\begin{figure*}[htbp]
\centering
\includegraphics[width=\linewidth]{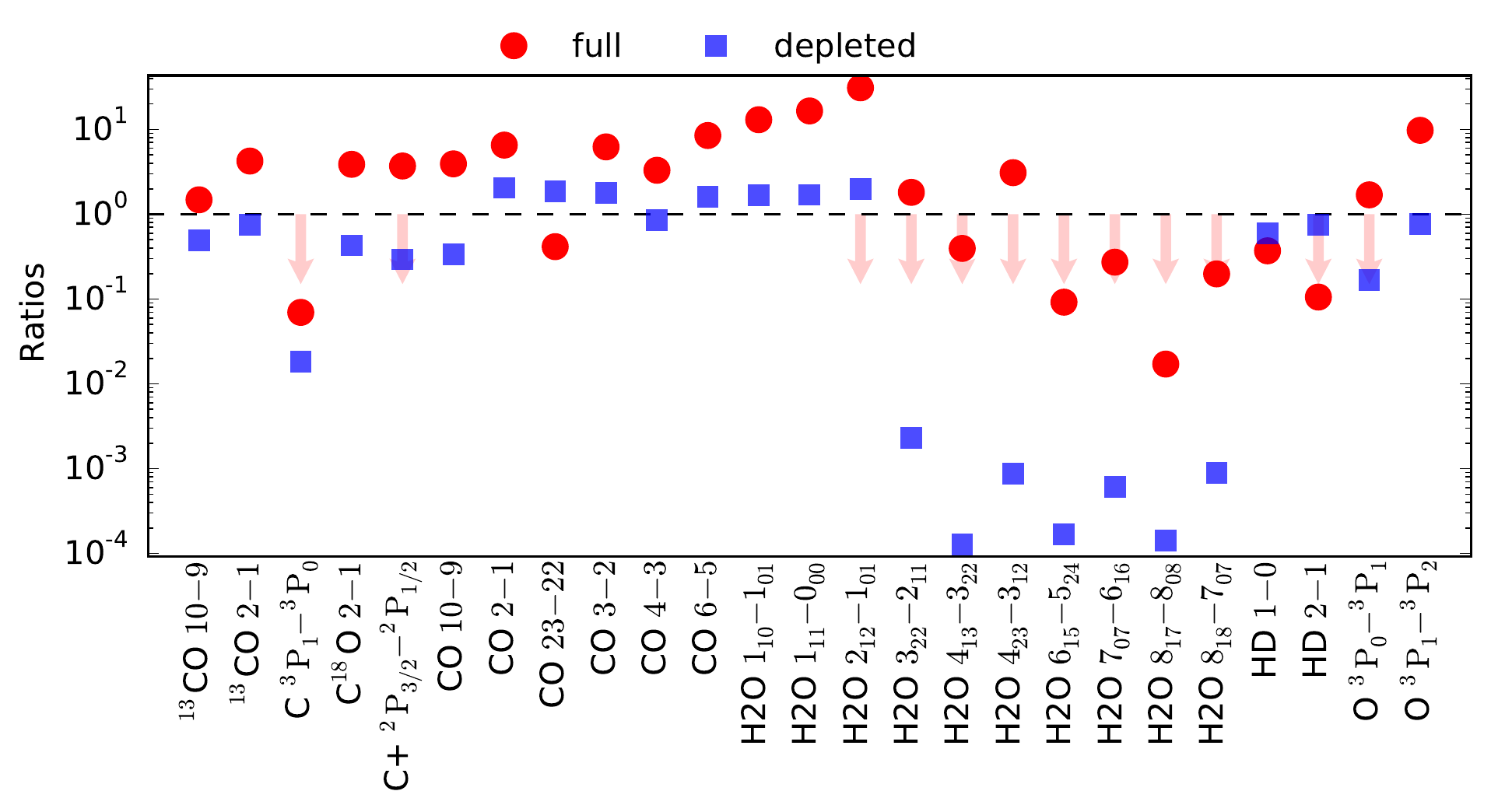}
\caption{Ratios between the modeled and observed intensities for a selection of
lines.  Red dot: with full oxygen and carbon abundances; blue rectangle:
with depleted oxygen and carbon abundances.  The arrows mean that the observed
values are upper limits.  The relative measurement error of the observational
data is usually much less than 50\%.  For the detected lines, a perfect fit
would land on the dashed line.}
\label{figCMP}
\end{figure*}

The code predict intensities for a suite of transitions of different species,
including \CII{}, \OI{}, H$_2$O, HD, and the CO isotopologues, which have been
detected by observation.  The resulting line intensities from the two models
are compared with the observed values in \reffig{figCMP} (for the numerical
values see \reftab{tabCMP}).  In agreement with previous studies
\citep{Thi2010,Kamp2013,Williams2014,Keane2014}, the model without elemental
depletion consistently over-predicts the line intensities of CO, water,
and a few other species for TW~Hya, while the model with depleted oxygen and
carbon abundances matches the observational data.  There are a few lines that
most clearly require oxygen and carbon to be depleted in the atmosphere of the
outer disk, including the \OI{} 63~$\mu$m, the CO and C$^{18}$O low-$J$ lines,
and all the water ground state lines.

\begin{table*}[htbp]
\centering
\begin{tabular}{llrrcccc}
\hline\hline
Molecule  & Line &  $\lambda$\phantom{xx}  &  E$_\text{up}$  &  Observed  & Full & Depleted  & Ref.      \\
\cline{5-7} &      &  ($\mu$m)   &  (K) &  \multicolumn{3}{c}{(10$^{-\text{18}}$~W~m$^{-2}$)}  &  \\
\hline
\ce{C+} & ${}^\text{2}\text{P}_\text{3/2}-{}^\text{2}\text{P}_\text{1/2}$     & 157.74  & 91   & $<$6.6     & 24.4   & 1.9  & 5  \\
\hline
\ce{C}  & ${}^\text{3}\text{P}_\text{1}-{}^\text{3}\text{P}_\text{0}$         & 609.14  & 24   & $<$0.4     & 0.03  & 0.01 & 7  \\
\hline
\multirow{2}{*}{\ce{O}} & $^\text{3}\text{P}_1-{}^\text{3}\text{P}_\text{2}$  &  63.18  & 228  &  36.5$\pm$12.1   & 356  &  31.3  & 5  \\
& $^\text{3}\text{P}_\text{0}-{}^\text{3}\text{P}_\text{1}$                   & 145.53  & 327 &   $<$5.5          & 9.3  & 1.1    & 5  \\
\hline
\multirow{5}{*}{\ce{CO}}
&  2$-$1                   & 1300.41 & 17   &  0.04   & 0.26     & 0.08  &  4 \\
&  3$-$2                   & 866.96  & 33   &  0.14   & 0.87     & 0.25  &  4 \\
&  4$-$3                   & 650.25  & 55   &  0.59   & 1.95     & 0.50  &  7 \\
&  6$-$5                   & 433.55  & 116  &  0.61   & 5.17     & 0.98  &  4 \\
& 10$-$9                   & 260.24  & 304  &  2.13   & 8.37     & 1.01  &  3 \\
& 23$-$22                  & 113.46  & 1524 &  4.4    & 1.82     & 8.22  &  3 \\
\hline
 $^\text{13}$CO & 2 $-$1   & 1300.41 & 17   &  0.02   &   0.08   & 0.02  &  2 \\
 $^\text{13}$CO & 10$-$9   & 260.24  & 304  &  0.36   &   0.53   & 0.19  &  3 \\
C$^\text{18}$O  & 2 $-$1   & 1300.41 & 17   &  0.006  &   0.02   & 0.003 &  2 \\
\hline
\ce{H2O} & -- & 21--35 & 800--3000  & \multicolumn{3}{c}{\multirow{2}{*}{See \reffig{figFitH2OOH}}}     &  6 \\
\ce{OH}  & -- & 21--35 & 400--4000  & & &  &  6 \\
\hline
\multirow{10}{*}{\ce{H2O}} 
& $\text{8}_\text{18}-\text{7}_\text{07}$ &  63.32 & 1071  &  $<$12.3       &  2.4  & 0.03   &  6 \\
& $\text{7}_\text{07}-\text{6}_\text{16}$ &  71.95 & 844   &  $<$12.7       &  3.5  & 0.02   &  6 \\
& $\text{8}_\text{17}-\text{8}_\text{08}$ &  72.03 & 1270  &  $<$3.5        &  0.06 & 0.008  &  6 \\
& $\text{4}_\text{23}-\text{3}_\text{12}$ &  78.74 & 432   &  $<$3.9        & 12.1  & 0.01   &  6 \\
& $\text{6}_\text{15}-\text{5}_\text{24}$ &  78.93 & 781   &  $<$5.4        &  0.5  & 0.003  &  6 \\
& $\text{3}_\text{22}-\text{2}_\text{11}$ &  89.99 & 297   &  $<$6.1        & 11.0  & 0.02   &  6 \\
& $\text{4}_\text{13}-\text{3}_\text{22}$ & 144.52 & 396   &  $<$0.6        &  0.2  & 0      &  6 \\
& $\text{2}_\text{12}-\text{1}_\text{01}$ & 179.53 & 114   &  $<$0.9        &  27.9 & 1.80   &  6 \\
& $\text{1}_\text{10}-\text{1}_\text{01}$ & 538.29 & 61    &  0.17$\pm$0.01 &  2.2  & 0.28   &  3 \\
& $\text{1}_\text{11}-\text{0}_\text{00}$ & 269.27 & 53    &  0.61$\pm$0.04 &  10.1 & 1.03   &  3 \\
\hline
\multirow{2}{*}{\ce{HD}} & 1$-$0          & 112.07 & 128   & 6.3$\pm$0.7    & 2.3   & 4.0    &  1 \\
                         & 2$-$1          & 56.23  & 384   & $<$8           & 0.8   & 6.0    &  1 \\
\hline
\end{tabular}
\tablerefs{
1. \citep{Bergin2013};
2. \citep{Favre2013};
3. \citep{Hogerheijde2011};
4. \citep{Qi2006};  synthesized beam corresponding to ~100 AU
5. \citep{Thi2010};
6. \citep{Zhang2013}.
7. \citep{Tsukagoshi2015}.
}
\caption{Comparison between modeled and observed integrated line fluxes.  The
``Full'' column is for the model with canonical carbon and oxygen abundances,
and the ``Depleted'' column is for the model with reduced carbon and oxygen
abundances.}
\label{tabCMP}
\end{table*}

\begin{figure}[htbp]
\centering
\includegraphics[width=\linewidth]{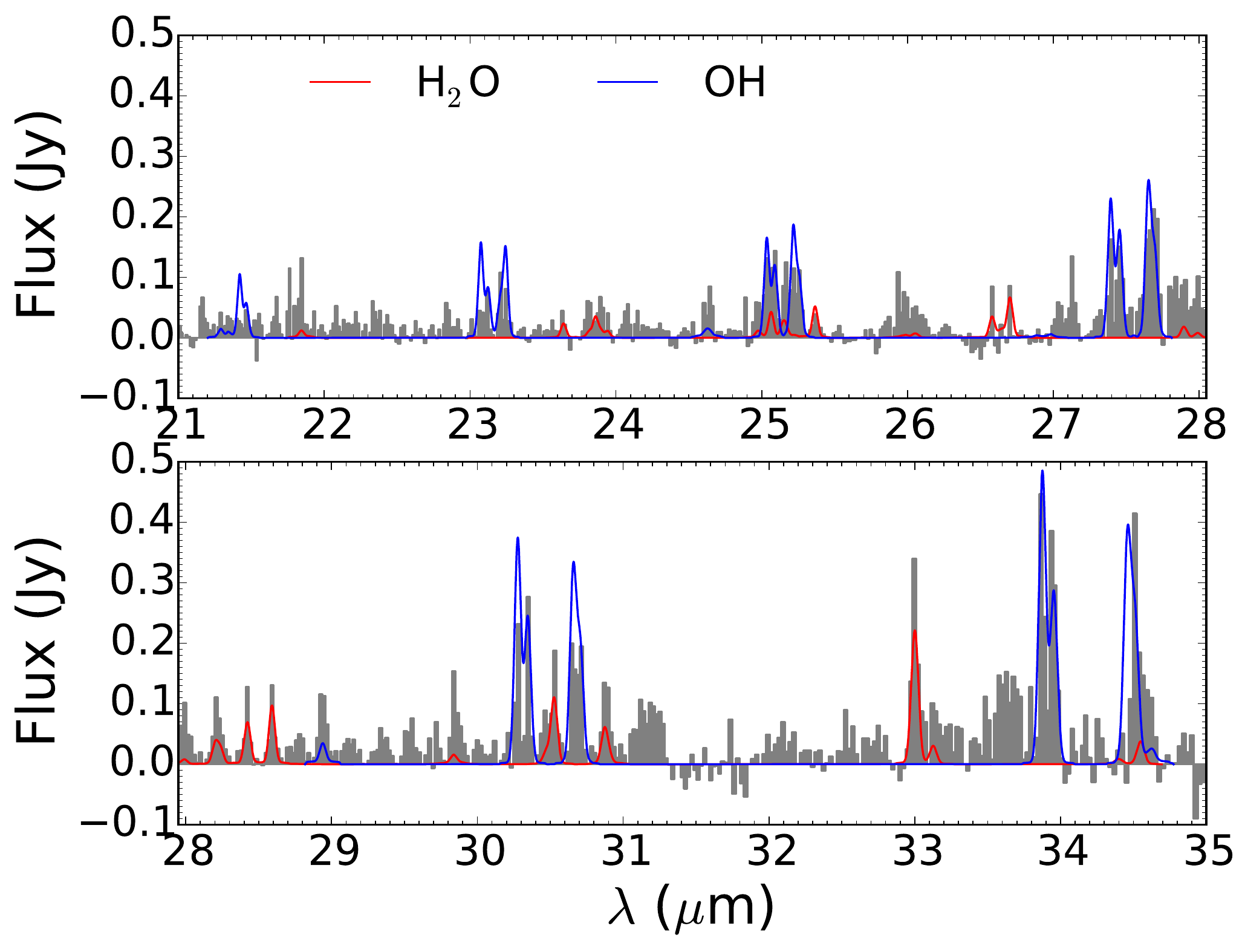}
\caption{Fitting to the \textsl{Spitzer/IRS} \ce{H2O} and OH lines.  The
\textsl{Spitzer} spectrum (in gray) is from \citet{Zhang2013}.}
\label{figFitH2OOH}
\end{figure}

In both models, the water $1_{10}-1_{01}$, $1_{11}-0_{00}$, and $2_{12}-1_{01}$
lines originate from $\gtrsim$20~AU, well beyond the water snow line.  Gas
phase water molecules in this region are mainly produced through
photodesorption of water ice \citep{Bergin2010, Hogerheijde2011, Kamp2013},
with their rates established by experiment \citep{Oberg2009a}.  A normal oxygen
abundance in the outer disk over-predicts their intensities by more than one
order of magnitude.  The situation is similar for the CO and C$^{18}$O low-$J$
lines (see also \citealt{Favre2013}).  The \OI{} and \CII{} lines mainly
originate from near the water snow line at $r\lesssim4$~AU in the model with
reduced oxygen and carbon abundances, coincident with where the water infrared
lines originate \citep{Zhang2013}.  With normal oxygen and carbon abundances,
additional contribution to the emission arises in gas out to 10 -- 20~AU, which
make them too strong, by a factor of 5{} in the case of \OI{}.  We note that a
deficit of elemental carbon in the disk atmosphere has been suggested for the
Herbig Be star HD~100546 by \cite{Bruderer2012}.


In the model with elemental depletion, we let the carbon and oxygen abundances
gradually return to normal when approaching the CO snow line ($\sim$20~AU) from
outside.  This is required to account for the observed high-$J$ and
rovibrational CO lines, as well as \OI{} and the numerous mid-infrared
transitions of water vapor (see \reffig{figFitH2OOH}).  This scenario resonates
with the recent result on the CO/\ce{H2} ratio, which is
$\sim1.6\times10^{-4}$ very close to the star in RW Aurigae~A
\citep{France2014}.  For the $^{13}$CO $10-9$ and CO $10-9$ and $23-22$
transitions, canonical carbon and oxygen abundances in the inner disk of the
depleted model are not enough to reproduce the data; these three lines are
under-predicted by about one order of magnitude.  Along our line of reasoning,
one intriguing possibility is that the carbon and oxygen abundances are
enhanced \citep{vanDishoeck2014}, rather than reduced, in the inner disk.  In
fact, if we increase the abundance of carbon and oxygen by a factor of 10 over
the ISM value within $\sim$15~AU (see the insets in panel (c) and (d) of
\reffig{fignTOC}), the CO $10-9$ and $23-22$ and the $^{13}$CO $10-9$ lines
would be in good agreement with observations, as shown in \reffig{figCMP},
while still keeping other lines consistent with observations.  Inward migration
and evaporation of icy bodies is predicted to cause such an enhancement of CO
and water vapor in the inner disk \citep{Cuzzi2004,Ciesla2006}.

\begin{figure}[htbp]
\centering
\includegraphics[width=\linewidth]{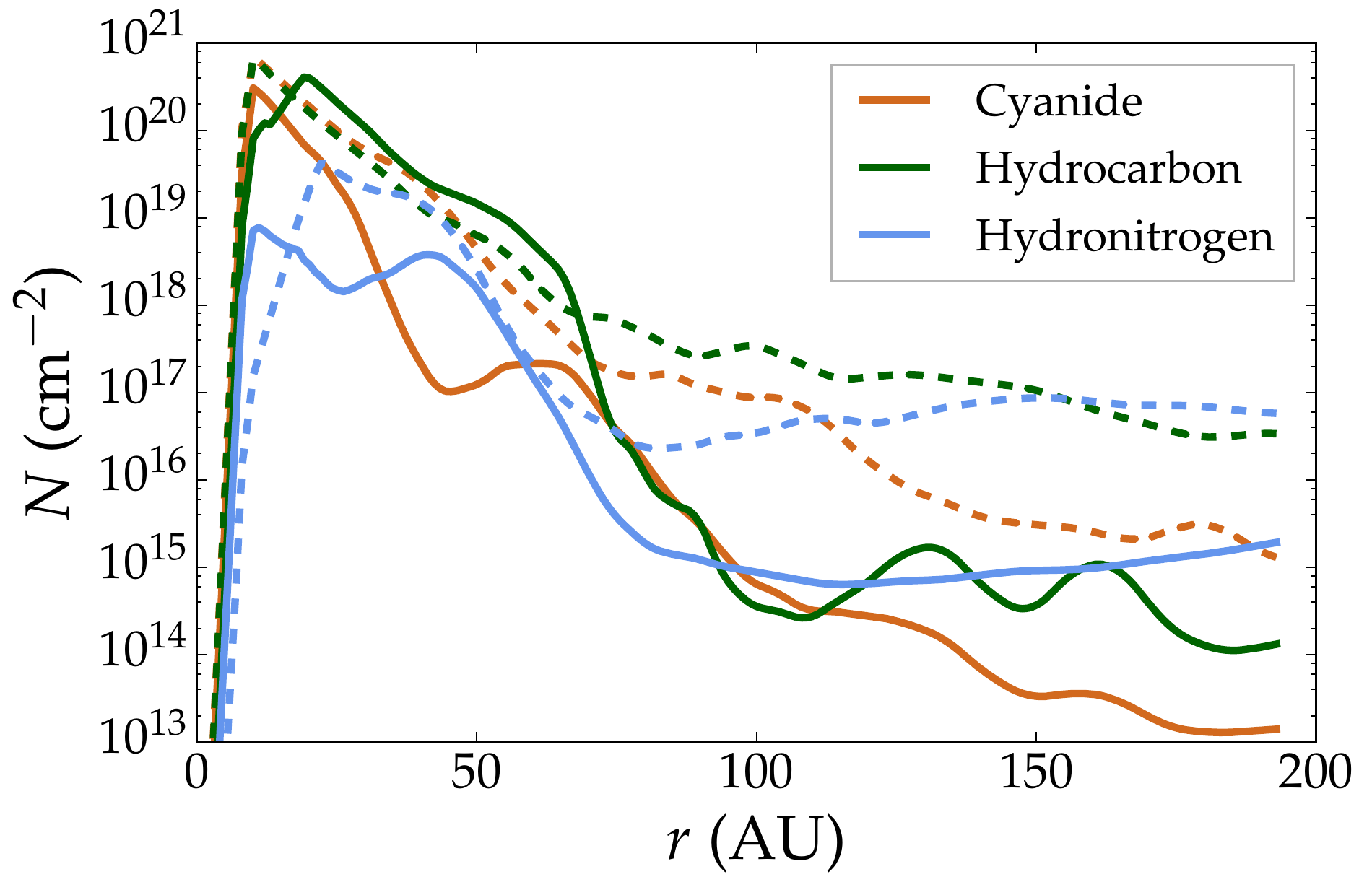}
\caption{Comparison between the column density of non-oxygen-bearing cyanide
(containing C-N bonds), hydrocarbon (containing C-H bonds), and hydronitrogen
(containing N-H bonds) from two models with full (solid lines) and reduced
(dashed lines) oxygen and carbon abundances.}
\label{figcmpCNH}
\end{figure}

In the above we have assumed that only oxygen and carbon are differentially
depleted in the outer disk atmosphere, and have left nitrogen intact.  This is
a reasonable assumption, considering that solar system ices as traced by comets
are well known to be nitrogen-poor \citep{Wychoff1991}.  One interesting outcome
of this assumption is that by depleting oxygen (and carbon, but to a lower
degree) over nearly the entire disk the gas will become rich in carbon
(particularly inside the CO snow line) and nitrogen-bearing molecules.  As an
example, \reffig{figcmpCNH} shows that the column densities of species containing C-N,
C-H, and N-H bonds can be increased by up to two orders of
magnitude in the depleted model.  This is in line with the recent
results on the enhanced abundances of C$_2$H and CN in evolved protoplanetary
disks \citep{Kastner2014}.

Depletion of carbon and oxygen in the outer disk also affect the HD emission.
This is because the cooling efficiency due to species bearing these two
elements are changed, so that the temperature in the outer disk becomes higher,
leading to a stronger HD emission.  For example, 80\% of the HD 112~$\mu$m
emission comes from within 80~AU in the model with no depletion, while with
depletion 60\% arises within the same radius (i.e. the outer disk contributes
more).  Though, as pointed out by \citet{Rollig2007}, the energy balance
calculation could be subject to considerable uncertainty in the tenuous
regions.

\section{Discussions}
\label{secDiscussions}

The depletion of oxygen and carbon relative to hydrogen solves the puzzle
regarding the gas mass of the TW~Hya disk.  The only known gas-loss mechanism
that preferentially removes species containing oxygen and/or carbon (mainly
H$_2$O and CO) rather than atomic or molecular hydrogen is ice mantle formation
on dust grains.  We emphasize that ice mantle formation and grain surface chemical reactions by themselves do not reduce
the local elemental abundances (gas + ice) in the atmosphere.
\citet{Reboussin2015} find that CO can be transformed into other species such as \ce{CO2} and \ce{CH4} on the dust grain surfaces (see also \citealt{Bergin2014}).
However, what we find is that if the
ice-coated grains stay in the upper atmosphere, desorption and dissociation by
UV photons will release \ce{H2O}, O, OH, CO, etc.\ to the gas phase, and the emission
from them will be much stronger than observations.
The key point is that, the
dust grains also grow through coagulation, and as they grow, they settle to the
disk midplane \citep{Dullemond2005a, Akimkin2013}, where they continue to grow
bigger and get assembled into planetesimals that may eventually grow to tens of
kilometers in size \citep{Stepinski1997}, possibly accompanied by inward drift
\citep{Laibe2012}.  The majority of the oxygen and carbon originally frozen
onto the dust grains in the atmosphere thus ends up inside solid bodies in the
midplane of the disk \citep{Bergin2010, Hogerheijde2011, Najita2013}.

A reverse process to dust settling described above must also be considered.
Dust grains in the upper layers can be replenished \citep{Kelling2011} by the
midplane reservoir through turbulent diffusion.  TW~Hya has relatively weak
turbulence \citep{Hughes2011}, which yields a ratio between the scale heights
of dust grains and gas of the order of 0.1 \citep{Cuzzi2006,Brauer2008}.  This
is consistent with the scale height of oxygen and carbon elements found in the
present work.

\citet{Birnstiel2014} find that the inward migration of dust grains combined
with gas drag produces a sharp edge in the dust distribution.  What we add to
this picture is that, the inward migration (together with growth and vertical
settling) of dust grains also removes species such as CO and \ce{H2O} from the
outer disk atmosphere, though emission of the leftover of these species that
remain in the outer region can still be significant.

The dust grains will eventually fall into the central star (may get
disintegrated on its way to the star), or become part of a planetesimal,
depending on whether they can overcome the fragmentation barrier
\citep{Johansen2008, Birnstiel2010} to grow large, and whether they can pass
the radial drift barrier to avoid falling into the star.   It is a theoretical
challenge to bypass the two barriers to maintain enough dust material in the
disk to form planets.  Empirically, there is evidence for
the early presence of large planetesimals for our solar system from
cosmochemical record of rocks.  It is known that parental bodies
($\gtrsim$50~km) of meteorites had already begun to form within a million years
of the condensation of calcium-aluminum-rich inclusions (CAIs)
\citep{Kleine2005, Qin2008, Kruijer2014}, and CAIs are generally
considered the oldest bodies of the solar system \citep{Amelin02}.  If systems
such as TW~Hya are similar to the solar system, then their planetesimals formed
at an early stage will be rich in carbon- and oxygen-bearing ices.

\acknowledgments F.D.{} and E.A.B.{} are supported by grant NNX12A193G from the NASA
      Origin of Solar Systems Program.
      We thank Ilse Cleeves and Fred Ciesla for discussions.


\end{document}